\documentclass[twocolumn]{aastex63}

\usepackage{array}
\usepackage{graphicx}
\usepackage{amsmath}

\shorttitle{Neutron stars formation \& retention through gas-assisted AICs}
\shortauthors{Perets, H. B.}

\graphicspath{{./}{figures/}}

\begin{document}
\title{Compact objects formation, retention and growth through accretion \\ onto gas-embedded white-dwarfs/neutron-stars in gas-enriched globular-clusters}

\author{Hagai B. Perets}
\affiliation{Physics department, Technion - Israel Institute of Technology, Technion city, Haifa 3200002, Israel}



\begin{abstract}
Observations of pulsars in globular clusters (GCs) give evidence that more $>10\%$-20$\%$ of neutron stars (NSs) ever formed in GCs were retained there. However, the velocity distribution of field pulsars peaks at 5-10 times the escape velocities of GCs. Consequently, only a small fraction of GC-NSs should have been retained, even accounting for low-velocity NSs formed through electron-capture supernovae. Thus, too few low-velocity NSs should have been retained, giving rise to the NS retention problem in GCs. Here we suggest a novel solution, in which the progenitors of most GC-NSs were ONe white-dwarfs (WDs) that accreted ambient intra-cluster gas and formed low-velocity NSs through accretion induced collapse (AIC). The existence of an early gas-enriched environment in GCs is supported by observations of GC multiple stellar populations. It is thought that 10s-100s of Myrs after the formation of the first generation of stars, and after ONe-WDs were already formed, GCs were replenished with gas which formed a second generation of stars. Accretion of such replenished gas onto the ONe-WDs catalyzed the AIC processes.
The number of AIC-formed NSs is then sufficient to explain the large number of NSs retained in GCs. Similar processes might also drive CO-WDs to produce type-Ia SNe or to merge and form NSs, and similarly drive NSs to AIC and mergers producing BHs. Moreover, the wide variety of gas-catalyzed binary mergers and explosive transients suggest to occur in the gas-rich environments of AGN disk could similarly, and even more efficiently, occur in second-generation gas in GCs.    
\end{abstract}

\section{Introduction}
For decades globular clusters (GCs) were thought to host simple stellar populations formed through a single star-formation episode. 
However, detailed photometric and spectroscopic studies collected over the last decade \citep[see e.g.][and references therein]{CBG09,Bastian2018} have shown that the vast majority of galactic GCs host multiple stellar populations with different light elements content. The exact origins of multiple stellar populations, have been extensively studied, but no clear solution has yet been found \citep[see][for a summary of the scenarios and their caveats]{Ren15, BL18, Gratton19}. The general consensus is that GCs experienced two or more star formation episodes, in which second population (2P) stars formed from processed gas lost by first population (1P) stars, and/or accreted external gas. Kinematics show that 2P stars are more centrally concentrated and were likely formed in the inner region of the GC where the second-generation gas likely accumulated.

While the exact source of the gas is debated, all suggested scenarios effectively require that tens up to hundreds of Mys after their formation, 1P stars had become embedded in a highly gas-rich environment that later produced the 2P stars. 
The evolution of stars, binaries and compact objects embedded in gas could therefore be significantly altered. Such processes were little studied in the context of the early second-generation gas-enriched environment \citep[][but see works by us and others on some aspects of such evolution]{ves+10,mac+12,lei+13,Leigh2014,Roupas2019}, but received much more attention in the context of gas-disks and AGNs environments near massive black holes  \citep[e.g.][and references therein]{art+93,BaruteauCuadraLin2011,McKernan2012,Stone2017,McKernan2018,Tagawa2020}.  

Here we explore the potential impact of the gas-enriched environment on the formation and evolution of neutron stars and their retention in clusters, and discuss other aspects of stellar and binary evolution of gas-embedded environments in clusters.

\cite{Pfa+02} pointed out the large numbers of pulsars and X-ray sources in GCs, suggesting large population of NSs exist in GCs.
They inferred that $>10-20\%$ of all NSs formed in GCs are retained there. Such high retention fractions are surprising and at odds with the observed velocity distribution of field pulsars; as GCs typically have low escape velocities of a few tens of km s$^{-1}$. Given the high natal kick velocities of single pulsars observed in the field \citep{hob+05,ver+17,Igo+21}, only a fraction of NSs should have been retained in GCs. Such inconsistency, sometimes termed the NS retention problem have been studied by various groups \cite[e.g.][]{pod05,iva+08}.  They and others suggested that a special population of low natal-kick velocity NSs is required in order to resolve the NS retention problem \citep{Pfa+02}. \cite{iva+08} proposed that NSs formed through electron-capture SN that receive low natal-ick velocities kicks could give rise to a few up to $10\%$ retention fraction for typical GCs, somewhat lower, but potentially comparable with the inferred retention fractions.
These results, however, were based on the assumption that electron-capture SNe progenitors extend over mass ranges of $6.85-7.57$ M$_\odot$ ($6.17-6.76$ M$_\odot$) for single stars with GC-like metallicity of Z = 0.005 (0.0005). However, \citet[and references therein]{wil+21} have recently shown that the mass range for electron-capture SN progenitors can not extend over a mass range larger than $\approx0.2$ M$_\odot$ given the observed velocities of isolated pulsars in the field. This would result in at least 3-5 times smaller numbers of single NSs formed through electron-capture SN, and hence similarly lower retention fractions, exacerbating the NS retention problem.  

Given the NS retention problem, but also irrespective of it, we suggest that the majority of NSs in GCs could have formed through efficient accretion of intra-cluster second-generation gas onto pre-existing Oxygen-Neon white dwarfs (ONe WDs). The accretion could drive such WDs to sufficiently large masses as to become unstable and collapse to form NSs through accretion induced collapse, AIC, \citep[e.g][]{can+76,nom+91}. AIC-formed NSs are thought to receive no, or very low natal kicks, allowing them to be retained in their host clusters.

In the following we first discuss the gas-enriched environment formed through delayed gas replenishemnt in GCs, and the expected conditions in such environments. We then discuss the accretion onto ONe WDs and its efficiency, and estimate the numbers of NSs that could form and be retained through AICs of ONe WDs. We then briefly explore the implications for accretion onto other compact objects and stars. Finally we  discuss and summarize the results.

\section{Early gas enrichment of globular clusters}
The total mass of second-generation gas in GCs is highly uncertain, but given reasonable assumptions on the relation between the gas and the observed second-generation stars in GCs one can provide an estimate the amount of replenished gas and its density. The mass in 2P stars ($\rm{M}_{\rm 2P}$) in clusters is of the order of half the GC mass \cite[with some correlation between the GC mass and the 2P fraction, with higher mass GCs having higher 2P fractions;][]{Bastian2018,MastrobuonoBattisiPerets2020}. 
Given the likely higher mass of newly formed GCs, compared with their inferred massed today, we consider the mass of 2P stars in a typical GC to be $\rm{M_{\rm 2P}}\approx 10^5 \rm{M}_\odot$ and given some efficiency of transforming gas into stars $\epsilon_g=0.3$ \cite[e.g. in models used by][]{Bekki2017}, we find the gas-mass to be $\rm{M_g}\approx 3.3\times10^5 M_\odot$. 

Observations suggest that the 2P stars concentrate in more compact configurations in the central part of GCs, consistent with simulation of 2P star formation in GCs, suggesting they mostly form and enclosed in the central pc of the cluster \citep{Bekki2017}, and likely in a disk configuration \citep{Bekki2010,MastrobuonoBattistiPerets2013,MastrobuonoBattisiPerets2016}.
Naturally, the densities in the cluster inner parts could be even higher, and the formation efficiency $\epsilon_g$ might be somewhat smaller or larger.

\cite{mac+12} adapted a typical density of $n=10^6$ gr cm$^{-3}$ \citep{pfa+09} following models by \cite{der+08} for gas replenishment by AGB stars. However, these models assume an exceedingly high star formation efficiency of 1, compared with e.g. Bekki's \citep{Bekki2017} models taking the efficiency to be 0.3, and considered a spherical configuration. It is likely that the 2P gas formed in a disk-like configuration, rather than a spherical distribution \cite[e.g]{Bekki2010,MastrobuonoBattistiPerets2013}. In order for stars to have formed the temperatures and hence the sound velocity of the gas should have been sufficiently low, down to $10$ km s$^-1$ or lower \cite{der+08,ves+10,Bekki2010}; if we adapt a sound speed of $c_=10$ km s$^{-1}$ \citep{der+08} we get a disk height of $H_d\approx (c_s/v)R_c$ where $v$ is the semi-Keplerian velocity in the cluster core and $R_c=1$ pc is the adapted size of the enriched gas region. We should note that \cite{pfa+09} explored external gas-accretion rather than gas from AGB stars and found far higher gas densities in the central region of up to $10^8-10^{10}$ for $10^6$ M$_\odot$ clusters.      

We can estimate the typical gas density by dividing the gas mass over the volume of a thick disk.
Here we will adapt a fiducial gas mass of $3.3\times10^5$ M$_\odot$ to get the atomic gas density in the disk (with size 1 pc and a scale height of H$_d\approx$0.23 pc) to be $n=1.9\times10^7$ cm$^{-3}$.  We briefly discuss the dependence on the density and its possible implications for the growth of ONe WDs.

The timescales for 2P gas replenishment and depletion are unknown.  Simulations \citep{Bekki2017} suggest gas replenishment can initiates on timescales of $\approx50$ Myrs, if arising from winds of AGB stars, leading to star-formation and gas depletion on the timescale of a few tens of Myrs, up to $\approx100-150$ Myrs. Observations and modeling of LMC young GCs showing multiple populations suggest potentially longer-replenishment timescales of a few hundred Myrs \citep{li+1}, with the gas originating from accretion of external, galactic disk gas by the GCs. The latter study does not provide clear constraints on the gas depletion timescales.

Here we adapt a simple model for gas depletion, assuming a constant gas density over a timescale $\tau_{acc}$ of 25 Myrs, and a fast (e.g. exponential) depletion, modeled for a simplicity as a strict immediate depletion at $\tau_{acc}$.  
\section{Accretion onto ONe white dwarfs}
\label{accreion}
The accretion rate onto ONe WDs, depends on the gas density in which they are embedded (discussed above) the relative velocity between the accretor and the gas, and the mode of accretion. 
In \cite{lei+13} we discussed the accretion rate on stars and the different processes involved. Here we briefly summarize the conclusions. 

The Bondi-Hoyle-Lyttleton (BHL) approximation for accretion onto gravitating objects in gas should be regarded as a strict upper limit to the
true accretion rate onto a compact object. It describes the rate at which
material becomes bound and is captured by an object at least temporarily. Feedback effects could later unbind some of the captured material giving rise to an effective lower mass accretion rate and growth of the accreting compact object (e.g. if the accretion is Eddington limited). Alternatively, or in addition, a WD accretor might not be able to efficiently grow in mass due to feedback from novae occurring on the accreting WDs, which could eject part or all of the accreted material.

Taking the fiducial gas-densities discussed above and following \citet{mac+12} and \cite{lei+13} the BHL accretion rate is given by re-scaling the formula in \citet{ho+03} to get:
\begin{equation}
\begin{gathered}
\label{eqn:mdot}
\dot{\rm M}_{\rm BHL} = 6.8\times10^{-8} \rm{M}_\odot \,{\rm yr}^{-1}\\\left( \frac{m}{1\rm {M}_\odot} \right) ^2 \left( \frac{n}{\rm 1.9\times10^7 cm
^{-3}} \right) \left( \frac{\sqrt{c_{\rm s}^2 + v_d^2}}{\rm 14 km s^{-1}} \right) ^{-3} \\
          = A m^2
\end{gathered}
\end{equation}
where $m$ is the accretor mass, $n$ is the gas density, $c_{\rm s}$ is the sound speed, and $v_d$ is the relative velocity between the gas and accretor for stars embedded in the disk, assumed to be comparable to the gas sound velocity, $v_d=10$ km s$^{-1}$. We assume atomic density of $n\approx1.9\times10^7$ cm$^{-3}$. The central temperatures were suggested to be a few thousand K, which translates into $c_{\rm s} \approx 10$ km s$^{-1}$ \citep{der+08}, though they might be lower for star-formation to occur \citep{Bekki2010}.  
These values represent the 2P gas-disk region, yielding
$A \approx 6.8\times10^{-8}$ M$^{-1}_\odot$ yr$^{-1}$ 
for ONe WD masses and velocities. The velocity dispersion dictated by the stellar and gas mass is of the order of $\approx43$ km s$^{-1}$, but would be only about a quarter of that for stars embedded in the disk. Hence, only about a quarter of the stars reside in the disk region (but possibly higher, given that GDF would lead to migration and capture of higher inclination stars into the disk; \citep[e,g][]{art+93}). 

\subsection{Accretion efficiency, growth and accretion-induced collapse}
The accretion efficiency of WDs, i.e. the fraction of accreted material retained and not ejected in novae through its accretion evolution depends on their mass, the accretion rate and their temperature. \cite{yar+05} calculated a grid of models for WD accretion. Their models explored higher and lower accretion rates than our fiducial rate. For accretion rates closet to ours of $10^{-7}$ M$_\odot$ yr$^{-1}$ they find that WDs in the mass range between 1.25-1.4 $M_\odot$ (relevant for ONe WDs), accrete at $5-30\%$ efficiency (with only one model M$_{\rm WD}$=1.25 M$_\odot$ with core temperature $\rm T=3\times10^7$ K, showing efficiency below 10$\%$). Magnetic WDs could potentially accumulate mass at lower accretion rates \cite{Abl+22}, but these are not considered here.
The final mass, M$_f$ of a WD with initial mass M$_i$ is of the order of 
\begin{equation}
    M_f = \frac{M_i}{1-M_iA\tau_{acc}f_{acc}}
\end{equation}
following \cite{lei+13} (and accounting for the retention efficiency, not used there for accretion on BHs). This would diverge for too long accretion times or initial masses, but significant accretion would deplete the gas \cite[e.g]{lei+13} before the scaling becomes unphysical.

Adapting a fiducial accretion time of $
\tau_{acc}=25$ Myrs and accretion efficiency of $f_{acc}=0.1$ (taking for simplicity a constant conservative low value between the highest and lowest retention efficiencies),
we find that WDs more massive than $\approx1.15$ M$_\odot$ will collapse to NSs, as they grow beyond $\approx1.38$ M$_\odot$ during the gas-depletion timescale.

In other words, for the plausible conditions of 2P star-formation in GCs,  ONe WDs would accrete sufficient mass as to grow up to the Chandrasekhar mass and transform to NSs through AIC.  

\section{Oxygen-Neon WD progenitors}
The formation and evolution of 1P ONe WDs depend on their progenitors stellar evolution. The evolution, in turn, depends on the metalicity of the stars. At lower metalicities, lower mass stars can give rise to ONe-WDs; the mass range of the progenitors then increases towards higher metalicities. Taking stellar evolution prescriptions \citep[using the SSE population synthesis code][]{hur+00} we find that at low Z=0.05Z$_\odot$ metalicity (Z$_\odot$=0.013), ONe (NS) progenitors are stars in the mass range 5.13-6.58 (compared with 6.58-18.44 for NSs) M$_\odot$  which form ONe-WDs after 60-100 Myrs. The progenitors mass range increases up to the range of 
6.2-7.9 (compared with 7.9-19.95 for NSs) M$_\odot$ for Solar metalicity, and formation timescales of the ONe WDs of 46-74 Myrs. In these models, ONe WDs have masses $>\approx1.2$ M$_\odot$, however, the lowest mass ONe WDs might have masses of $1.1$ M$_\odot$, in which case the lower progenitor mass, could be $\approx0.4$ M$_\odot$ lower than assumed here (and progenitor lifetimes could extend up to 90 Myrs even for Solar metalicity). 

Given the above discussion of the  replenishment timescales, we find these are generally consistent with 1P ONe WDs already formed before, or at most during the main gas-replenishment epoch, allowing them to interact with, and accrete 2P gas.  
\section{NS retention fraction}
The NS retention fraction is typically defined as the fraction of NSs retained out of the NS formed in a cluster. However, in previously considered models AIC-formed NSs contributed a negligible fraction of the NSs. In particular the inferred retention fractions of 10-20$\%$ calculated by \cite{Pfa+02} referred to NSs formed from massive stars, and not through AIC of WDs. Consequently, in the following we consider the fraction of retained NSs out of the NSs formed from massive stars, for consistent comparison.    

Given the progenitor mass range for ONe WDs, we can find the relative numbers of initially formed ONe-WDs and of NSs assuming typical initial mass-functions \citep[e.g. Miller-Scalo IMF][]{mil79}. Given efficient accretion allowing the ONe-WDs to form NS through AIC, we find that the number of AIC formed NSs is $\approx61-65\%$ of the NS formed from massive stars. Assuming, like \cite{iva+08} that AIC-formed NSs get low-velocity kicks (10 times smaller than regular NSs, i.e. assuming a Maxewllian velocity distribution with a 10-times lower mean than inferred by \cite{hob+05}, about $85\%$ of such AIC-formed formed low velocity NSs were retained in our fiducial cluster. 

If we extend ONe WD masses to as low as 1.15 $M_\odot$ (such WDs would grow to 1.38 M$_\odot$ and follow an AIC), these fractions could extend to 1.5 larger. As mentioned above, only stars embedded in the disk will efficiently accrete. Assuming the relative fraction of stars in the disk goes like the volumetric ratio of the disk volume of the central pc, we get only $\approx17\%$ of the stars reside in the disk. Overall we then find a retention fraction 9-14$\%$ with these assumptions. Due to GDF, it is likely that a larger fraction of the stars could migrate into the disk \citep[as suggested for AGN disks; e.g.][]{art+93}, raising these fractions even higher. On the other hand, newly formed second generation stars would dilute these fractions by up to a factor of two.  
Taken together, we expect $\approx5-15\%$ retention fractions, given the many uncertainties; these are generally consistent with the inferred NS retention fractions in GCs \citep{Pfa+02}.  


\section{Discussion}
\subsection{Accretion, growth, AIC and explosion of gas-embedded WDs}

We find that AIC of ONe-WDs due to accretion of intracluster gas could potentially produce the majority of NSs retained in GCs. AIC is likely to occur when an ONe WD approaches $\approx1.38$ M$_\odot$, although the exact mass could depend on spin and and/or other properties. Models suggest that $\approx0.1$ M$_\odot$ of material is ejected during the collapse, and the AIC-formed NSs should therefore have typical initial masses close to $1.3$ M$_\odot$, although the newly born NS might continue to accrete, if still embedded in intracluster gas \citep{lei+13}, or if they later accrete from a binary companion.

The BHL accretion rate on lower mass CO WDs is lower than the accretion rate on the more massive ONe WDs. However, their mass-retention rate is higher \citep{yar+05}, allowing them to to grow in mass. Lower mass CO WDs could therefore potentially grow, even up to the point they reach the Chandrasekhar mass, and explode as type Ia SNe. However, the first CO WDs form only $\approx100$ Myrs after the GC formation, and therefore, depending on the timing of the gas-replenishment into the GC, only the most massive CO might form in time to accrete. Nevertheless, for gas replenishment occurring on longer time-scales of 300 Myrs, CO WDs as low as $\approx0.95$ M$_\odot$ and down to $\approx0.84$ (for $Z=0.05Z_{\odot}$ and Solar metalicity, respectively)  could form,  accrete, and potentially explode as type Ia SNe, though the gas-depletion timescale required might be 1.2-1.5 times longer than 25 Myrs.    

Accretion onto gas-embedded WDs in AGNs was first discussed (to the best of our knowledge) by \cite{ost83}, who suggested such accretion could lead to type Ia SNe. More recent studies by \cite{gri+21,zhu+21} explore the feedback and appearance of explosive thermonuclear explosion in AGN disks, suggesting that accretion can help drive such explosions, but with little discussion of the actual accretion rates. However given the comparable gas densities, one can indeed expect to attain effective growth of WDs at least in some regions of AGN disks. Our proposed mechanism for low-kick NS formation in gas-embedded GCs, should therefore be also applicable to the AIC production of NSs in AGN disks, or in gas-replenished nuclear clusters which could experience several epochs of star formation.   

\cite{mac+12} first explored the role of accretion onto WDs in gas-embedded GCs. In particular, they focused on the role of novae arising from the accretion process and its potential pollution of the 2P gas and stars. Their paper is in many ways complementary to our proposed accretion processes. They focus on the ejected mass through novae rather than on the growth of the WDs. The main difference is in the assumptions regarding the accretion and retention rates. 

\cite{mac+12} assume a gas density of 2P gas which is a few times lower than in our models, while also suggesting a much longer lifetime of the gas in the GC before its dispersal through star-formation and feedback processes.
These differences could have significant implications.  In particular, the transition between efficient retention of material by accretion WDs and inefficient retention \citep{yar+05} effectively occurs in-between the accretion rate taken by \cite{mac+12} and the accretion rate we discuss above. Nevertheless, given the uncertainty in the amount of 2P gas, it is quite plausible that a range of gas densities could exist in different GCs and even in different regions in the same GC.
It is therefore possible for both \cite{mac+12} and our scenario to operate at different clusters, or even in the same cluster at different regions/times, allowing for both inefficient accretion on some WDs and the pollution of GCs by novae, and the growth and AIC of other WDs and the production and retention of low-velocity AIC-formed NSs.  

\subsection{Accretion, growth, AIC and explosion of gas-embedded {\emph NSs} and {\emph BHs}}
Although we focused on accretion, growth and AIC of WDs, similar processes could potentially operate for NSs \citep{col+99}. \cite{per+21} explored the possibility of accretion onto gas-embedded NSs in AGN disks. Most of their results would therefore be generally applicable to NSs embedded in 2P gas in GCs, including spin-up of NSs and the AIC formation of low-mass BHs from accretion NSs.  However, we point out some important differences. First, NSs formed through core-collapse SNe could be retained in nuclear clusters near MBHs, where the escape velocities are far higher than in GCs; or migrate there through slow mass-segregation processes \cite[e.g.][and references therein]{aha+16}. Therefore NSs could reside in AGN disks and accrete there. Such NSs, however would mostly be expelled from GCs, as discussed above, and would not be able to participate in such processes. Nevertheless, as mentioned earlier, if 2P gas depletion timescales are sufficiently long (e.g. 100-300 Myrs), NSs formed and retained through the proposed accretion and AIC of WDs discussed here could then have sufficient time to further grow after their formation and follow the evolution discussed by \cite{per+21}. In other words, our proposed WD accretion and AIC model are also a prerequisite for wide-scale NS growth and AIC of NSs into BH to occur in GCs.

Stochastic isotropic accretion onto NSs might only lead to growth, without spinning-up NSs, but if the NSs are embedded in a more coherent disk-like structure in GCs \citep[e.g.][and references therein]{Bekki2010,MastrobuonoBattistiPerets2013,MastrobuonoBattisiPerets2016}, accretion could be more coherent in term of angular momentum accretion, and lead not only to growth, but also to spin-up of NSs.

The growth and evolution of BHs through accretion in gas-embedded GCs is also possible, as we discussed in \cite{lei+13}, potentially giving rise to the formation of very high mass BHs, even beyond the pair-instability mass cut-off \citep{Roupas2019}. Again, the level of spin-up from such accretion would depend on the structure of the gaseous component and the orbits of the stars in comparison with the gas.   

\subsection{Accretion onto main-sequence stars}
The role of environmental cluster gas accretion on non compact-objects stars was first discussed by \cite{art+93}, with more recent studies by \cite{dav+20} and \cite{Can+21} (and references therein). As already discussed by us in \cite{lei+13}, more massive stars can accrete significantly and grow. We point out that similar processes would occur in second-generation gas in GCs. In particular, less massive stars (e.g. $<2-5$ M$_\odot$ stars) which could still be on the main-sequence (MS) during the gas replenishment could sufficiently grow, as to effectively change the mass-function of 1P stars in GCs  and even end their life in core-collapse SN and form NSs/BHs, i.e. produce NSs from stars that would otherwise have only formed WDs. As with NSs, stochastic isotropic accretion onto stars might lead to growth, without spinning them up. If stars are embedded in a more coherent disk-like structure in GCs as discussed above, stars could be spun up and thereby affect their stellar evolution and the core-collapse supernova of the more massive stars, and the formed NSs/BHs, and in turn the possibility of producing an electron-capture supernova producing low-velocity natal kick or long-GRBs following core-collapse of fast-rotating massive stars. 

Accretion onto stars could potentially deplete the gas as discussed by us in \cite{lei+13}, however, it is quite possible that once grown to become very massive feedback (e.g. winds and radiation) may quench further accretion. 
The general aspects of stellar evolution of, and accretion on MS stars embedded in 2P gas in GCs is beyond the scope of this paper and will be discussed elsewhere. We do note that if the 2P gas originates from winds of evolved stars, suggested to explain the chemical differences between 1P and 2P clusters, it could pollute the envelopes of MS stars. Given the very low-mass of the outer convective envelope of low-mass stars, even a slight pollution (e.g. accretion of only $0.01$ M$_\odot$) onto 1P stars would make them appear as 2P stars, masquerading their true 1P nature, at least until other processes, such as thermohaline mixing could dilute the low-mass convective envelope from the polluted material. One might therefore expect a much larger fractions of 2P stars during the early evolution of GCs, on the thermohaline mixing timescales. We leave discussion on this somewhat disconnected topic to further studies.      
\subsection{Gas-catalyzed compact binary mergers}
The role of gas-catalyzed mergers of compact objects and their implications was discussed by us in the context of binary planetesimlas embedded in protoplanetary disks \citep{per+11,gri+16} and later in the context of compact object binaries embedded in AGN disks \citep{McKernan2012}, where the latter have been extensively studied since then \citep[e.g.][and references therein]{Stone2017,McKernan2018,Roupas2019,Tagawa2020}. The 2P gas environment in GCs could provide similar and even more favorable conditions for gas-catalyzed mergers (e.g. larger fraction of hard-binaries, lower velocity dispersions, and far larger numbers of massive clusters compared with nuclear clusters in AGNs). Here we focus on accretion processes on WDs and their implications for NS formation and production; a dedicated study of binary mergers in such environments is explored in a companion paper (Rozner and Perets, 2021, in prep.). However, we should note here that mergers of ONe-WDs with CO or other ONe-WDs could also give rise to a {\emph merger} induced collapse of the remnant merged object, which, similar to the AIC case, would likely involve no, or very little natal kicks, hence providing another channel for NSs formation and retention in gas-embedded GCs. Such process could produce more massive NSs, with typical masses of the combined masses of the merging WDs, $\approx1.8-2.4$ M$_\odot$, given that only massive WDs should exist at the time of 2P gas replenishment.  

\subsection{NS fractions dependence on cluster properties}
Larger gas reservoirs would give rise to larger gas densities and faster and more efficient accretion and growth. If the fractions of 2P stars could correspond to the size of the gas reservoir, then one would expect some correlation between the NS fractions and the 2P star fractions. The NSs (effective known mostly from millisecond pulsars) are known to correlate with the collisional parameter of the cluster, which in turn correlates with the cluster mass. The 2P stellar population is also known to correlate with the cluster mass \cite[e.g][and references therein]{MastrobuonoBattisiPerets2020}. Consequently, a correlation between the NS population and the 2P population is expected, but would be difficult to disentangle from the formation processes of 2P stars and the collisional dynamics that play a role in the formation of millisecond pulsars used to asses the NS population. That being said, with sufficient future statistics and better understanding of 2P star-formation one might be able to differentiate between the various dependencies in the future.  

\subsection{Potential caveats and uncertainties}
Most of the main potential caveats and uncertainties involved in the proposed scenario have been discussed in previous sections. These can be briefly summarized as follows:
\begin{itemize}
    \item Potentially large uncertainties exist in regard to the gas densities and hence the accretion rates and mass retention and growth rate of WD. Too low densities would not allow for efficient WD growth, as also discussed in \cite{mac+12}.
    \item The accretion rate was calculated assuming a BHL accretion; the accretion rate could, however be more limited due to feedback effects (e.g. Eddington limited), leading to lower accretion rates; see further discussion of this in \cite{lei+13}. 
    \item The timescales for gas replenishment and the timescale for the gas depletion are not known. The former would affect which compact objects will be able to accrete (longer timescales allow for lower mass compact objects to form before 2P gas replenishment ensues). The latter would affect the timescale for the accretion and hence the overall mass growth of compact objects. Furthermore, accretion onto stars and compact objects might lead to faster depletion of gas and/or give rise to feedback effects, e.g. from supernovae of newly grown massive stars.
\end{itemize}

\section{Summary}
In this paper we explored the effect of accretion of second generation intra-cluster gas onto WDs in globular clusters (or young mssive clusters), and its implication for producing and retaining NSs in GCs. The observations of multiple stellar populations in GCs, gave rise to various models for gas replenishment in GCs which could form two or more stellar populations separated by tens up to hundreds of Myrs. The evolution and interaction of 1P stars and compact objects (formed during the first epoch of star-formation) with the large gas reservoirs that later form the 2P stars could lead to accretion of replenished intra-cluster gas onto the stars and compact-objects. 

Considering plausible gas densities and timescales for the gas replenishment and depletion (motivated by observations and simulations of multiple population GCs) we find that massive WDs, and in particular ONe-WDs would already form and then accrete up to a few$\times$0.1 M$_\odot$ of material, allowing them to reach the Chandrasekhar limit and form low-velocity NSs through accretion induced collapse. The numbers of such intra-cluster gas catalyzed AIC-formed NS could explain the origin of most NSs in GCs and their retention there, alleviating potential challenges raised to previously proposed solutions to the GC NS retention problem. Although we considered plausible gas densities and accretion rates, we should note that lower gas-densities and/or lower accretion rates might give rise to inefficient accretion and growth of WDs, which could then limit the proposed channel for AIC formation of NSs. 

Similar gas accretion processes onto NSs and BHs could give rise to the production of BHs both in the low mass BH gap-mass regime ($<3-5$ M$_\odot$ from AIC and gas-catalyzed mergers of NSs, as well as very massive BHs in the upper (pair instability) BH mass-gap region. Accretion onto main-sequence 1P stars can pollute low-mass stars and change the mass-function of intermediate stars.

Massive reservoirs of second generation intra-cluster gas have strong observational support from the existence of multiple stellar populations in GCs. Our findings \cite[][and Rozner \& Perets, in prep.]{lei+13,Leigh2014} and the current discussion on accretion onto compact objects and the novel channel for NSs formation and retention, suggest a major revision is needed in the modeling of the early evolution of GCs. The early gas-embedded GC evolution would give rise to major differences in both the overall dynamics of GCs and the stellar and compact-objects populations in these environments. These in turn would directly propagate into the modeling of various stellar exotica, X-ray binaries, millisecond pulsars and gravitational-waves sources expected to form in GCs. 

I would like to thank Noam Soker for important comments that improved this manuscript. HBP acknowledges support for this project from the European Union's Horizon 2020 research and innovation program under grant agreement No 865932-ERC-SNeX.

\bibliographystyle{aasjournal}



\end{document}